\documentclass[fleqn,10pt,a4paper]{article}
\usepackage{amsmath,amssymb,bm}
\usepackage{cite,epsfig,color,graphicx}
\title{Exploring Born-Infeld electrodynamics using plasmas}
\author{DA Burton\thanks{Department of Physics, Lancaster University, UK 
and the Cockcroft Institute, UK}
\and
RMGM Trines\thanks{Rutherford Appleton Laboratory, Chilton, Didcot, UK
  and Department of Physics, Lancaster University, UK.}
\and
TJ Walton\footnotemark[1]
\and 
H Wen\footnotemark[1]
\and
}
\begin{document}
\maketitle
%%%%%%%%%
\begin{abstract}
The behaviour of large amplitude electrostatic waves in cold plasma is
investigated in the context of Born-Infeld electrodynamics. The
equations of motion for the plasma are established using an
unconstrained action principle. The maximum amplitude and frequency of
a large amplitude electrostatic wave are determined, and a lower bound
on the wavelength is established. The maximum electric field is found
to be the same as that on a point electron at rest. 
\end{abstract}
\section{Introduction}
%%%%%%%%%
%\thispagestyle{fancy}
%\rhead{{\bf Cockcroft-08-29}}
Recent years have seen renewed interest in non-linear
electrodynamics~\cite{gibbons:2001, ferraro:2007,
  dereli:2010}, in particular Born-Infeld theory~\cite{born:1934}, and its
implications. Unlike their Maxwell counterparts, the Born-Infeld
field equations are \emph{fundamentally} non-linear; they are
non-linear even in the vacuum. Born and Infeld introduced
their theory~\cite{born:1934} in order to ameliorate the singular
self-energy of a point charge. Furthermore, it was discovered that among the family of non-linear
generalizations of Maxwell electrodynamics, Born-Infeld theory possesses
a number of highly attractive features; in particular, like the vacuum Maxwell equations,
the vacuum Born-Infeld equations exhibit zero birefringence and its
solutions have exceptional causal behaviour~\cite{boillat:1970,
  plebanski:1968}. Furthermore, Born-Infeld theory shares a number of
properties with the low energy dynamics of strings and
branes~\cite{fradkin:1985}.

Some of the most extreme conditions ever encountered in a terrestrial
laboratory are created when high-power laser pulses interact with
matter. The laser pulse immediately vaporizes the matter to form an
intense laser-plasma providing novel avenues for generating intense
bursts of coherent electromagnetic radiation for a wide range of
applications in biological and material science~\cite{schlenvoigt:2008}. Furthermore,
laser-plasmas permit controllable investigation of matter in extreme
conditions that only occur naturally away from the Earth.
It is expected that the next generation of ultra-intense
lasers will, for the first time, allow controllable access to regimes
where a host of different quantum electrodynamic phenomena will be
evident~\cite{marklund:2010} and theories of non-linear electrodynamics will be
central to their study. However, it is conceivable that Born-Infeld
electrodynamics may play a role before it is necessary to invoke quantum
electrodynamics~\cite{dereli:2010}.

Motivated by recent analyses of the propagation of Born-Infeld
electromagnetic waves in waveguides~\cite{ferraro:2007} and in background uniform magnetic
fields~\cite{dereli:2010}, this article is addresses aspects of the non-linear
electrodynamics of laser-plasmas. 

A sufficiently short and intense laser pulse propagating through a
plasma may create a travelling longitudinal plasma wave whose velocity
is approximately the same as the laser pulse's group
velocity. However, it is not possible to sustain arbitrarily large electric fields; substantial
numbers of plasma electrons become trapped in the wave and are
accelerated, which dampens the wave (the wave `breaks'). Early theoretical investigation
of non-linear plasma waves was undertaken by Akhiezer and
Polovin~\cite{akhiezer:1956}, and later expounded by
Dawson~\cite{dawson:1959} in the context of wave-breaking. 

Wave-breaking is a fundamentally non-linear phenomenon, and it is
natural to explore the properties of Born-Infeld plasmas from this
perspective. For simplicity, in the present article we adopt the
\emph{cold} plasma model for the electron fluid and generalize
well-known results~\cite{akhiezer:1956, dawson:1959}, for the cold Maxwell
plasma, to the context of non-linear electrodynamics. In particular,
we obtain an exact expression for the maximum electric field of
electrostatic oscillations in a cold Born-Infeld plasma and analyse
the wavelength and frequency of those oscillations. There are numerous
discussions of electrostatic waves in a warm Maxwell plasma; for
example, see~\cite{burton:2010, trines:2006}. 

We employ the Einstein summation convention throughout this article.
Latin indices $a,b,c$ run over $0,1,2,3$ and units are used in which the speed of light
$c=1$ and the permittivity of the vacuum $\varepsilon_0=1$.

Let $(x^a)$ be an inertial coordinate system on Minkowski spacetime
$({\cal M},g)$ where $x^0$ is the proper time of observers at fixed
Cartesian coordinates $(x^1,x^2,x^3)$ in the laboratory. The metric tensor
$g$ has the form
\begin{equation}
g = \eta_{ab} dx^a \otimes dx^b
\end{equation}
with
\begin{equation}
\eta_{ab} =
\begin{cases}
&-1\text{   if $a=b=0$},\\
&1\text{    if $a=b\neq 0$},\\
&0\text{    if $a \neq b$}
\end{cases}
\end{equation}
and the Hodge map $\star$ is induced from the $4$-form $\star 1$ on ${\cal
  M}$ where
\begin{equation}
\star 1 = dx^0 \wedge dx^1 \wedge dx^2 \wedge dx^3.
\end{equation}

The plasma electrons are represented
as a cold relativistic fluid; their worldlines are trajectories of
the unit normalized future-pointing timelike $4$-vector field $V$ on
${\cal M}$ and the $0$-form $n$ is their proper number density.
We are interested in the evolution of a plasma over timescales during
which the motion of the ions is negligible in comparison with the
motion of the electrons, and we assume that the ions are at rest and
distributed homogeneously in the laboratory frame. Their worldlines
are trajectories of the vector field $N_{\rm ion} = n_{\rm
  ion}\partial/\partial x^0$ on ${\cal M}$ where $n_{\rm ion}$ is the
ion number density (a positive constant) in the laboratory frame.

Maxwell's equations may be written covariantly as
\begin{equation}
\label{maxwell}
dF = 0,\qquad d\star G = - q n\star\widetilde{V} - q_{\rm ion} \star\widetilde{N_{\rm ion}}
\end{equation}
where $q<0$ is the charge on the electron and $q_{\rm ion} = Z|q|$ is the charge on
an ion with $Z$ the multiplicity of the ionization. The $1$-forms
$\widetilde{V}$ and $\widetilde{N_{\rm ion}}$ are the metric duals of
the vector fields 
$V$ and $N_{\rm ion}$ respectively, i.e. the $1$-form $\widetilde{V}$
satisfies $\widetilde{V}(U)=g(V,U)$ for all
vector fields $U$ on ${\cal M}$.
The Maxwell $2$-form $F$ encodes the electric field
${\bf E}$ and the magnetic induction ${\bf B}$; the excitation $2$-form $G$ encodes the electric
displacement ${\bf D}$ and the magnetic field ${\bf H}$. In
Maxwell electrodynamics $G=F$ in the vacuum; however, in Born-Infeld
electrodynamics (\ref{maxwell}) is retained but the constitutive relations for $({\bf D},{\bf
  H})$ in terms of $({\bf E},{\bf B})$ are non-linear in the
vacuum. In Born-Infeld electrodynamics the constitutive relation for
$G$ in terms of $F$ is  
\begin{equation}
\label{BI_constitutive}
G = \frac{1}{\sqrt{1-\kappa^2 X - \kappa^4 Y^2/4}}(F - \kappa^2 Y
\star F)
\end{equation}
where the invariants $X$ and $Y$ are
\begin{equation}
\label{invariants}
X = \star (F\wedge \star F),\qquad Y = \star (F\wedge F)
\end{equation}
and $\kappa$ is a new constant of nature.

To motivate the field equations used herein, we begin
section~\ref{section:action_principle} by developing them using an
unconstrained action principle~\cite{kijowski:1998}. We will briefly review how
(\ref{BI_constitutive}) arises and show that $V$ satisfies the field
equations
\begin{equation}
\label{lorentz}
\nabla_V \widetilde{V} = \frac{q}{m}\iota_V F,\qquad g(V,V)=-1
\end{equation}
for a relativistic cold electron fluid, 
where $q \iota_V F$ is the Lorentz $4$-force acting on the electron
fluid, $\iota_V$ is the interior product with respect to $V$, $m$ is
the electron rest mass and $\nabla$ is the Levi-Civita
connection on ${\cal M}$. This will be followed in
section~\ref{section:non-linear_oscillations} by an
exploration of the properties of large amplitude relativistic
electrostatic oscillations.
\section{Action principle}
%%%%%%%%%
\label{section:action_principle}
A succinct action principle for establishing (\ref{maxwell},
\ref{lorentz}) is formulated using a {\it three}-dimensional  
manifold ${\cal B}$ (often called a `material'~\cite{maugin:1993},
or `body', manifold), where each point in ${\cal B}$ corresponds to
an integral curve of $V$.

The independent variables of the action introduced below are the
electromagnetic $1$-form $A$ on ${\cal M}$ and a submersion $f$ from ${\cal M}$ to ${\cal B}$,
\begin{eqnarray}
    f: {\cal M} &\longrightarrow& {\cal B} \\
          x^{a} &\longmapsto& \xi^{A} = f^{A}(x).
\end{eqnarray}
The map $f$ identifies an integral curve of $V$ with its corresponding
point in ${\cal B}$, where upper case Latin indices $A,B = 1,2,3$.
The inverse image $f^{-1}(P)$ of the point $P\in{\cal B}$ is the
worldline of an idealized `material' particle in the electron fluid
and $V$ may be determined from $f$ as the unique timelike future-pointing vector field satisfying
\begin{equation}
df^A(V) = 0
\end{equation}
and
\begin{equation}
g(V,V) = -1.
\end{equation}

In the present approach, material properties of the electron fluid are
encoded in tensors and forms on ${\cal B}$, and fields on
${\cal M}$ are induced from material tensors on ${\cal B}$ using $f$. In particular, the
number of electrons $N[\Sigma]$ occupying the spacelike hypersurface
$\Sigma$ may be induced from a $3$-form $\Omega$ on ${\cal B}$ as
\begin{equation}
N[\Sigma] = \int_{\Sigma} f^*\Omega
\end{equation}
where $f^*$ denotes the pull-back map, induced from $f$, on forms on ${\cal B}$ to
forms on ${\cal U}$. The electron number current $3$-form $j$,
\begin{equation}
\label{definition_j}
j=f^*\Omega
\end{equation}
is identically closed,
\begin{equation}
dj = 0,
\end{equation}
where $d$ is the exterior derivative, which follows because $\Omega$ is a
top form on ${\cal B}$ and $df^* = f^*d$.

A variational principle yielding (\ref{lorentz}) for $V$ and
(\ref{maxwell}) for the potential $1$-form $A$, where $F=dA$, employs the action
functional $S$ where
\begin{equation}
\label{action}
S[A,f] = \int_{\cal M} [{\cal L}_\text{EM}(X,Y)\star 1 - m\sqrt{j\cdot
  j}\star 1 -
  q A\wedge j - q_\text{ion} A\wedge j_\text{ion})]. 
\end{equation}
The $0$-form ${\cal L}_\text{EM}(X,Y)$ is a local function of the
invariants $X$ and $Y$ only. For standard Maxwell
electrodynamics ${\cal L}_\text{EM} = X/2$, whereas for Born-Infeld
electrodynamics ${\cal L}_\text{EM} = {\cal L}_\text{BI}$ and
\begin{equation}
\label{definition_LBI}
{\cal L}_\text{BI}(X,Y) = \frac{1}{\kappa^2}(1-\sqrt{1-\kappa^2 X -
  \kappa^4 Y^2/4}).
\end{equation}
The $3$-form $j_\text{ion}=n_\text{ion} dx^1\,\wedge dx^2\, \wedge dx^3$ is the background ion number
current and the $0$-form $j\cdot j$ is the square of the magnitude of $j$,
\begin{equation}
j \cdot j = \star^{-1}(j\wedge\star j)
\end{equation}
where $\star^{-1}$ is the inverse of the Hodge map $\star$.

The first term ${\cal L}_\text{EM}(X,Y)\star 1$ in the Lagrangian $4$-form in
(\ref{action}) depends only on $A$ and
$g$, the second term $m\sqrt{j\cdot j}\star 1$ depends only on $f$ and
$g$, the third term $q A\wedge j$ couples $f$ and $A$ and the fourth
term $q_\text{ion} A\wedge j_\text{ion}$ depends only on $A$.

We will show in the following that the field equations (\ref{maxwell},
\ref{lorentz}) for $n$, $V$ and $A$ are
recovered from those for $f$ and $A$ by introducing the electron proper number density $n$ and
$4$-velocity field $V$ as 
\begin{align}
\label{definition_n}
&n = \sqrt{j\cdot j},\\
\label{definition_V}
&\widetilde{V} = \frac{\star^{-1} j}{n}.
\end{align}
\subsection{Non-linear generalization of the Maxwell equations}
%%%%%%%%%%%
Equations for the electromagnetic field $F$ arise upon seeking stationary
variations of $S$ with respect to $A$ :
\begin{equation}
\delta_A S = \int_{{\cal M}} \bigg[\bigg(\frac{\partial {\cal
      L}_\text{EM}}{\partial X} \delta_A X + \frac{\partial {\cal
      L}_\text{EM}}{\partial Y} \delta_A Y\bigg)\star 1 - \delta A \wedge
      (qj + q_\text{ion} j_\text{ion})\bigg]
\end{equation}
where $\delta_A \alpha$ denotes the variation of a form $\alpha$ with respect to
 $A$. Using (\ref{invariants}) and $\star\star\alpha = -\alpha$ where
 $\alpha$ is a $4$-form on ${\cal M}$, it follows
\begin{equation}
\label{invariants_starred}
X\star 1 = - F\wedge \star F,\qquad Y\star 1 = - F\wedge F
\end{equation}
and
\begin{equation}
\delta_A X \star 1 = \delta_A (X\star 1) = -2d\delta A \wedge \star
F,\qquad \delta_A Y \star 1 = \delta_A
(Y\star 1) = -2d\delta A \wedge F.
\end{equation}
Hence
\begin{equation}
\delta_A S = \int_{{\cal M}} \delta A\wedge \bigg[-2 d\bigg(\frac{\partial{\cal
      L}_\text{EM}}{\partial X} \star F +\frac{\partial{\cal
      L}_\text{EM}}{\partial Y} F\bigg) -
      (qj + q_\text{ion} j_\text{ion})\bigg] 
\end{equation}
where the variation $\delta A$ is chosen to have compact support on
${\cal M}$, and an integration by parts has been used in the final
step.  By demanding that the action is stationary under all such
variations we obtain
\begin{equation}
d\star G = - qj - q_\text{ion} j_\text{ion}
\end{equation}
where
\begin{equation}
\label{def_star_G}
\star G = 2 \bigg(\frac{\partial{\cal
      L}_\text{EM}}{\partial X} \star F +\frac{\partial{\cal
      L}_\text{EM}}{\partial Y} F\bigg).
\end{equation}
The field equations (\ref{maxwell}) are obtained by introducing $n$ and
$V$ using (\ref{definition_n}) and (\ref{definition_V}), and the ion
number $4$-current $N_\text{ion}$ as
\begin{equation}
\widetilde{N_\text{ion}} = \star^{-1}j_\text{ion}
\end{equation}
and noting $dF=0$ since $F$ is an exact $2$-form.
Equation (\ref{BI_constitutive}) immediately follows from
(\ref{definition_LBI}, \ref{def_star_G}) with ${\cal L}_\text{EM} =
{\cal L}_\text{BI}$ and using $\star\star\beta = -\beta$ where
 $\beta$ is a $2$-form.
\subsection{Field equations for the electron fluid}
%%%%%%%%%%%
The field equations (\ref{lorentz}) describing the electron fluid are obtained by seeking stationary
variations of $S$ with respect to $f$. Using (\ref{definition_n}) it
follows
\begin{equation}
\delta_f \sqrt{j\cdot j} = \frac{1}{\sqrt{j\cdot j}}[\star^{-1}(\delta_f j\wedge\star j)].
\end{equation}
To proceed further we express $j=f^* \Omega$ explicitly in
terms of the components $\{f^A\}$ of $f$.
\begin{equation}
\label{variation_j_wrt_f}
\begin{split}
\delta_f j &= \frac{1}{3!} \bigg(\frac{\partial\Omega_{ABC}}{\partial
  \xi^D}\circ f\bigg) \,
\delta f^D\,df^A\wedge df^B\wedge df^C +
  \frac{1}{2!}(\Omega_{ABC}\circ f)\, d\delta f^A \wedge df^B \wedge df^C\\
&= \iota_W dj + d\iota_W j
\end{split}
\end{equation}
where $\Omega = \frac{1}{3!}\Omega_{ABC}(\xi)\,d\xi^A\wedge
d\xi^B\wedge d\xi^C$ and
\begin{equation}
j = f^*\Omega = \frac{1}{3!}(\Omega_{ABC}\circ f)\,df^A\wedge df^B\wedge df^C
\end{equation}
have been used. The vector field $W$ is
\begin{equation}
W = \delta f^A W_A
\end{equation}
where $\{W_A\}$ is a basis for $V$-orthogonal vector fields on ${\cal
  M}$, and the frame $\{V, W_A \}$ and coframe $\{-\widetilde{V},
  df^A\}$ are naturally dual, i.e.
\begin{equation}
df^A(W_B) = \delta^A_B,\quad \widetilde{V}(W_A) = 0,\quad df^A(V) =
0,\quad \widetilde{V}(V) = -1
\end{equation}
where $\delta^A_B$ is the Kronecker delta.
Since $j$ is a closed $3$-form it follows
\begin{equation}
\delta_f j = d\iota_W j
\end{equation}
using (\ref{variation_j_wrt_f}) and so
\begin{equation}
\begin{split}
\delta_f(\sqrt{j\cdot j}\star 1) &= \frac{1}{\sqrt{j\cdot j}} \delta_f j \wedge\star j\\
&= d\iota_W j \wedge \frac{1}{\sqrt{j\cdot j}}\star j.
\end{split}
\end{equation}

Varying (\ref{action}) with respect to $f$ yields 
\begin{equation}
\delta_f S = -\int_{\cal M} \bigg[\frac{m}{\sqrt{j\cdot j}} \delta_f j \wedge\star j + q A
  \wedge \delta_f j\bigg]
\end{equation}
where the variation $\delta f^A$ is chosen to have compact support on
${\cal M}$. An integration by parts yields
\begin{equation}
\delta_f S = \int_{\cal M} \iota_W j \wedge d(\frac{m}{\sqrt{j\cdot j}}\star j - q A)
\end{equation}
and by demanding that $S$ is stationary under all suitable variations $\delta f^A$ we
obtain
\begin{equation}
\label{field_equation_for_j}
\iota_{W_A} j \wedge d(\frac{m}{\sqrt{j\cdot j}}\star j - q A) = 0.
\end{equation}

Eliminating $j$ in favour of $n$ and $V$ using (\ref{definition_n},
\ref{definition_V}) yields
\begin{equation}
\iota_{W_A}\iota_V (m d\widetilde{V} - q F) = 0
\end{equation}
and since $\{V,W_A\}$ is a frame on ${\cal M}$ it follows
\begin{equation}
\label{lorentz_iota}
\iota_V d\widetilde{V} = \frac{q}{m} \iota_V F.
\end{equation}
Since $g(V,V)$ is constant it may be shown $\iota_V d\widetilde{V} = \nabla_V \widetilde{V}$
(see, for example,~\cite{burton:2003}) and (\ref{lorentz}) follows
immediately.
\section{Non-linear electrostatic oscillations}
%%%%%%%%
\label{section:non-linear_oscillations}
The remainder of this article focuses on properties of
large-amplitude longitudinal electrostatic waves propagating parallel to the $x^3$-axis with phase
velocity $v$  (with $0 < v <1$) in the laboratory frame. We introduce
the pair $\{e^1,e^2\}$
\begin{equation}
e^1= vdx^3 - dx^0, \qquad e^2=dx^3 - vdx^0
\end{equation}
and note that the orthonormal coframe $\{\gamma e^1,\gamma e^2,
dx^1, dx^2\}$ is adapted to observers moving at velocity $v$ along
$x^3$ (i.e observers in the `wave frame'). We seek a $4$-velocity $V$ of the form
\begin{equation}
\label{V_ansatz}
\widetilde{V} = \mu(\zeta) e^1 + \psi(\zeta) e^2
\end{equation}
where $\zeta = x^3 - vx^0$ is the wave's phase and we have adopted the so-called
`quasi-static approximation'; the
pointwise dependence of $\mu$ and
$\psi$ is on $\zeta$ only. Using $g(V,V) = -1$ the component $\psi$ is
found as
\begin{equation}
\label{negative_psi}
\psi = -\sqrt{\mu^2-\gamma^2}
\end{equation}
where the Lorentz factor $\gamma=1/\sqrt{1-v^2}$ and the sign of $\psi$ is
chosen to ensure that the velocity $\gamma e^2 (V)$ of the electron
fluid in the wave frame is non-positive. Thus, in the laboratory frame
the speed of the electrons is slower than the phase speed of the wave,
except at wave-breaking where the electrons catch the wave. 

In the present analysis we assume that the electromagnetic field is
due entirely to the electron fluid and ion background. The magnetic
field vanishes and the only
non-zero component of the electric field is in the $x^3$
direction; it follows that the Maxwell $2$-form $F$ is
\begin{equation}
\label{def_E}
F = E dx^0\wedge dx^3
\end{equation}
where $E$ is the $x^3$ component of the electric field.

Since $\widetilde{V}$ and $F$ are elements of the subspace of forms on
${\cal M}$
generated by $\{dx^0, dx^3\}$ it follows 
\begin{equation}
\label{V_wd_dV_vanishes}
\widetilde{V}\wedge (d\widetilde{V} - \frac{q}{m}F) = 0
\end{equation}
and, using (\ref{lorentz_iota}, \ref{V_wd_dV_vanishes}) with $g(V,V)=-1$, we obtain 
\begin{equation}
\label{dV_F}
d\widetilde{V} = \frac{q}{m} F
\end{equation}
and (\ref{V_ansatz}, \ref{def_E}, \ref{dV_F}) yield
\begin{equation}
\label{electric_field}
E = \frac{1}{\gamma^2} \frac{m}{q} \frac{d\mu}{d\zeta}.
\end{equation}

Since $d\zeta \wedge F = 0$ and $dF = 0$ it follows 
\begin{equation}
\label{closure_dLdYF}
d\biggl(\frac{\partial{\cal L}_\text{EM}}{\partial Y} F\bigg) = 0
\end{equation}
and inserting (\ref{electric_field}, \ref{V_ansatz}) into
(\ref{maxwell}) yields
\begin{equation}
\frac{d}{d\zeta}\bigg(2\frac{\partial{\cal L}_\text{EM}}{\partial
  X}\bigg|_{Y=0}\frac{d\mu}{d\zeta}\bigg) = \frac{q^2}{m} (n\mu - Z n_\text{ion}),\qquad n = \frac{Z
  n_\text{ion} \gamma^2 v}{\sqrt{\mu^2-\gamma^2}}  
\end{equation}
where
\begin{equation}
Y=0
\end{equation}
follows from (\ref{invariants}) and (\ref{closure_dLdYF}) has been used. Hence 
\begin{equation}
\label{ODE_for_mu}
\frac{d}{d\zeta}\bigg(2\frac{\partial{\cal L}_\text{EM}}{\partial X}\bigg|_{Y=0}\frac{d\mu}{d\zeta}\bigg) =
\frac{q^2}{m} Z n_\text{ion}
\gamma^4\bigg(\frac{v\mu}{\sqrt{\mu^2-\gamma^2}} - 1\bigg).
\end{equation}
The product of (\ref{ODE_for_mu}) and $d\mu/d\zeta$ may be written
\begin{equation}
\label{diff_of_first_integral}
\frac{d}{d\zeta}\bigg[\bigg(2\frac{\partial{\cal L}_\text{EM}}{\partial
  X} X - {\cal L}_\text{EM}\bigg)\bigg|_{Y=0} - m
  Z n_\text{ion}(v\sqrt{\mu^2-\gamma^2} - \mu) \bigg] = 0
\end{equation}
using
\begin{equation}
X=E^2,
\end{equation}
which follows from (\ref{invariants}). 
\subsection{Maximum amplitude oscillation}
\begin{figure}
\begin{center}
\scalebox{1.0}{\includegraphics{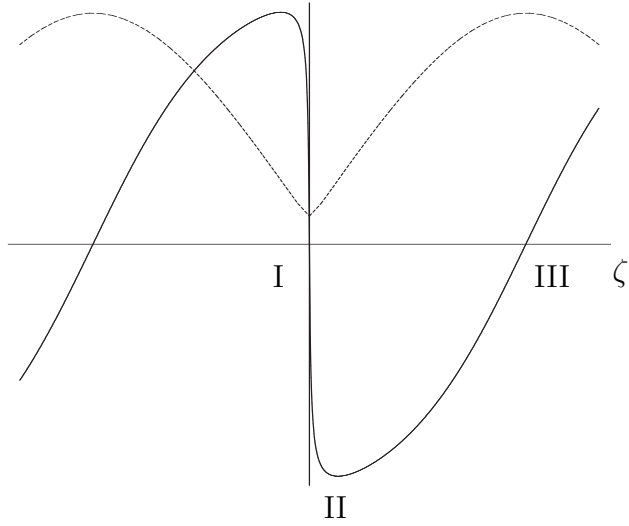}}
\caption{\label{fig:mu_vs_zeta_plots} The dashed curve shows $\mu$ versus
$\zeta$ and the solid curve shows $E$ versus $\zeta$ (not to
  scale). Points of intersection of $E$ with the $\zeta$-axis are
  labeled ${\rm I}$ and ${\rm III}$, and ${\rm II}$ is a turning point
of $E$. Using (\ref{electric_field}) and $q<0$ it follows
$d\mu/d\zeta$ and $E$ are of opposite sign.}
\end{center}
\end{figure}

%%%%%%%%%%%
The square root in the right-hand side of (\ref{ODE_for_mu}) places a
lower bound on $\mu$. Suppose that over an oscillation the component $\mu$ of
$\widetilde{V}$ attains its lowest possible value $\mu_\text{I} =
\gamma$ at $\zeta=\zeta_\text{I}$ (see
figure~\ref{fig:mu_vs_zeta_plots}). Since $\mu_\text{I}$ is a turning point of $\mu$ it follows
$E(\zeta_\text{I})=0$. At $\zeta=\zeta_\text{II}$ the electric field
$E(\zeta_\text{II}) = - E_\text{max}$ where $E_\text{max}$ is the
amplitude of the largest possible oscillation (see
figure~\ref{fig:mu_vs_zeta_plots}).
Integration of (\ref{diff_of_first_integral}) over
$[\zeta_\text{I},\zeta_\text{II}]$ yields
\begin{equation}
\label{Emax2}
\bigg(2\frac{\partial{\cal L}_\text{EM}}{\partial
  X} X - {\cal L}_\text{EM}\bigg)\bigg|_{
\begin{subarray}{l}
X=E^2_\text{max}\\
Y=0
\end{subarray}
}
+ {\cal L}_\text{EM}\bigg|_{
\begin{subarray}{l}
X=0\\
Y=0
\end{subarray}
}
= m Z n_\text{ion}(\gamma - 1)
\end{equation}

For a standard cold Maxwell plasma ${\cal L}_\text{EM} =
X/2$, which inserted into (\ref{Emax2}) yields the well-known
wave-breaking limit due to Akhiezer and Polovin~\cite{akhiezer:1956}
\begin{align}
\label{Emax_AP}
E^\text{AP}_\text{max} &= \sqrt{2m Z n_\text{ion} (\gamma - 1)}\\
&= \frac{m\omega_p c}{|q|}\sqrt{2(\gamma - 1)}
\end{align}
for a multiply-ionized relativistic cold plasma, where
\begin{equation}
\omega_p = \sqrt{\frac{q^2 Z n_\text{ion}}{m\varepsilon_0}}
\end{equation}
is the plasma frequency and the speed of light $c$ and permittivity of
the vacuum $\varepsilon_0$ have been restored.

Using ${\cal L}_\text{EM} = {\cal L}_\text{BI}$ and
(\ref{definition_LBI}, \ref{Emax2}, \ref{Emax_AP}) it follows
\begin{align}
\label{Emax_BI}
E^{\text{BI}}_\text{max}
= \frac{1}{\kappa}\sqrt{1 - [\kappa^2 (E^\text{AP}_\text{max})^2/2 + 1]^{-2}}
\end{align}
for a cold Born-Infeld plasma and $\lim_{\kappa\rightarrow
  0}E^{\text{BI}}_\text{max} = E^\text{AP}_\text{max}$ as expected.

The choice $\kappa = \varepsilon_0
r_0^2/|q|$ made by Born and Infeld, where $r_0$ is the classical radius
of the electron, leads to $\kappa\sim 10^{-22}\,{\rm m}/{\rm
  V}$. However, quantum electrodynamic vacuum effects are expected to become
important at electric field strengths $\sim 10^{18}\,{\rm V}/{\rm m}$, and if
Born-Infeld theory plays a role at the classical level it follows
$E^{\text{BI}}_\text{max} \lesssim 10^{18}\,{\rm V}/{\rm m}$
and $\kappa E^{\text{BI}}_\text{max}\ll 1$. The next generation of
laser systems (such as Extreme Light Infrastructure (ELI)~\cite{eli:2010} and
High Power Laser Energy Research system (HiPER)~\cite{hiper:2010}), are expected to offer
intensities $\sim 10^{25}\,{\rm W}/{\rm cm}^2$ corresponding to
electric field strengths $\sim 10^{16}{\rm V}/{\rm m}$.  

Although in practice $\kappa E^{\text{BI}}_\text{max}\ll 1$ it is
worth noting that although the limit of $E^{\text{AP}}_\text{max}$ as
$v\rightarrow c$ does not exist, using (\ref{Emax_BI}) it follows
\begin{equation}
\label{limit_EBI}
\underset{v\rightarrow c}{\lim} E^{\text{BI}}_\text{max} = \frac{1}{\kappa}
\end{equation}
which is the value of the electric field of a static point electron
evaluated at the location of the point electron in Born-Infeld
electrodynamics~\cite{born:1934}.
% Finally, the non-relativistic limit of
% $E^{\text{BI}}_\text{max}$ is Dawson's expression for the maximum
% electric field of a cold Maxwell plasma,
%\begin{equation}
%\label{Emax_dawson}
%E^{\text{BI}}_\text{max} = \frac{m\omega_p v}{|q|}[1  + {\cal O}(v^2/c^2)],
%\end{equation}
%and it is clear that exploration of the $\kappa$ dependence of
%$E^{\text{BI}}_\text{max}$ for $v\ll c$ requires the relativistic
%corrections in (\ref{Emax_dawson}).
\subsection{Period and frequency of maximum amplitude oscillations}
%%%%%%%%%%%  
The period $\lambda$ of the
maximum amplitude oscillation of a cold Born-Infeld plasma is
obtained from the solution to (\ref{ODE_for_mu}) with ${\cal L}_{\rm
  EM} = {\cal L}_{\rm BI}$ and the initial conditions $\mu(\zeta_{\rm
  I}) = \gamma$ and $d\mu/d\zeta|_{\zeta = \zeta_{\rm I}} = 0$. The particular first
integral of (\ref{ODE_for_mu}) satisfying the initial conditions on
$\mu$ may be written
\begin{equation}
\label{dmudzeta2}
\bigg(\frac{d\mu}{d\zeta}\bigg)^2 = \frac{q^2\gamma^4}{m^2\kappa^2}\bigg(1-[\kappa^2
     m Z n_{\rm ion}(v\sqrt{\mu^2 - \gamma^2} - \mu + \gamma) + 1]^{-2}\bigg)
\end{equation}
and, using (\ref{dmudzeta2}), consideration of the stationary points
of $\mu$ yields $\gamma\le \mu \le \gamma^3(1+v^2)$. Furthermore, since $\mu(\zeta_{\rm I}) = \gamma$ it follows
$\frac{d\mu}{d\zeta} > 0$ for $\zeta_{\rm I} < \zeta < \zeta_{\rm III}$ where $\mu(\zeta_{\rm
  III}) = \gamma^3(1+v^2)$ is the maximum value of $\mu$ (see
figure~\ref{fig:mu_vs_zeta_plots}). Thus, using (\ref{dmudzeta2}) the
  period $\lambda$ of the maximum amplitude oscillation is
\begin{equation}
\label{lambda_integral}
\begin{split}
\lambda &= 2(\zeta_{\rm III} - \zeta_{\rm I})\\
&= \frac{2}{\omega_p \gamma^2} \int\limits^{\gamma^3(1+v^2)}_\gamma
\hat{\kappa} \frac{1}{\sqrt{1-[\hat{\kappa}^2
     (v\sqrt{\mu^2 - \gamma^2} - \mu + \gamma) +
     1]^{-2}}}\, d\mu
\end{split}
\end{equation}
where $\hat{\kappa} = \kappa m \omega_p/|q|$.

A lower bound on $\lambda$ may be determined by noting $0\le (v\sqrt{\mu^2-\gamma^2} -
\mu + \gamma) \le \gamma - 1$ for $\gamma\le \mu \le
\gamma^3(1+v^2)$. It follows
\begin{equation}
\frac{1}{\sqrt{1-[\hat{\kappa}^2
     (v\sqrt{\mu^2 - \gamma^2} - \mu + \gamma) +
     1]^{-2}}} \ge \frac{1}{\sqrt{1-[\hat{\kappa}^2
     (\gamma - 1) +
     1]^{-2}}}
\end{equation}
and thus
\begin{equation}
\label{lower_bound_lambda}
\begin{split}
\lambda &> \frac{2}{\omega_p \gamma^2} \int\limits^{\gamma^3(1+v^2)}_\gamma
\hat{\kappa} \frac{1}{\sqrt{1-[\hat{\kappa}^2
     (\gamma - 1) +
     1]^{-2}}}\, d\mu\\
&= \frac{4m\gamma v^2}{|q| E^{\text{BI}}_\text{max}}. 
\end{split}
\end{equation}
Hence, using (\ref{limit_EBI}, \ref{lower_bound_lambda}) it follows
$\lambda$ diverges as least as fast as $\gamma$ in the limit
$v\rightarrow c$. Thus, although the Lorentz force on an electron
trapped in a maximum amplitude wave is finite, the estimate $4m\gamma
v^2/|q|$ of the work done on the electron over a wave period diverges
in the limit $v\rightarrow c$
and, in principle, it is possible to accelerate electrons to
arbitrarily high energies in the present classical
theory. Furthermore, consideration of the electromagnetic
stress-energy-momentum tensor ${\cal T}_\text{EM}$ arising from metric
variations of the first term in (\ref{action})~\cite{dereli:2010}
yields an electromagnetic mass-energy density $\varrho_\text{EM}$ (in
the lab frame) that diverges in the limit $v\rightarrow c$ for a wave on the verge of
breaking. The tensor ${\cal T}_\text{EM}$ satisfies
\begin{equation}
\label{EM_stress_tensor}
{\cal T}_\text{EM}(T,U) = \frac{\partial{\cal
      L}_\text{EM}}{\partial X}\iota_T\star^{-1}\bigg(\iota_U
      F\wedge\star F - F\wedge \iota_U \star F\bigg) + \bigg({\cal
      L}_\text{EM} - X\frac{\partial{\cal L}_\text{EM}}{\partial X} -
      Y\frac{\partial{\cal L}_\text{EM}}{\partial Y}\bigg)\,g(T,U)
\end{equation}
where $T,U$ are arbitrary vectors, and the $0$-form
$\varrho_\text{EM}$ is
\begin{equation}
\label{T00}
\varrho_\text{EM} = {\cal
  T}_\text{EM}(\partial/\partial x^0,\partial/\partial x^0).
\end{equation}
Using (\ref{def_E}, \ref{EM_stress_tensor},
\ref{T00}) it follows
\begin{equation}
\label{EM_energy_density}
\varrho_\text{EM} = \bigg(2X\frac{\partial{\cal L}_\text{EM}}{\partial
  X} - {\cal L}_\text{EM}\bigg)\bigg|_{Y=0}
\end{equation}
and the Born-Infeld energy density $\varrho_\text{BI}$ obtained using
(\ref{EM_energy_density}, \ref{definition_LBI}),
\begin{equation}
\varrho_\text{BI} = \frac{1}{\kappa^2}\frac{1}{\sqrt{1-\kappa^2
    X}}(1-\sqrt{1-\kappa^2 X}),
\end{equation}
diverges in the limit $E\rightarrow 1/\kappa$.
 
The behaviour of (\ref{lambda_integral}) for $\hat{\kappa}^2
\gamma \ll 1$ and  $\gamma \gg 1$ may be extracted by changing the integration
variable to $\chi = \mu/\gamma^3$. Using (\ref{lambda_integral}) it follows
\begin{equation}
\label{lambda_integral_approx}
\lambda = \frac{2\gamma}{\omega_p}\int\limits^{1+v^2}_{\gamma^{-2}}
\frac{1}{\sqrt{2 I}} \bigg[ 1+\frac{3}{4}\hat{\kappa}^2 I + {\cal
      O}(\hat{\kappa}^4 I^2) \bigg]\, d\chi
\end{equation}
where
\begin{equation}
\label{definition_I}
\begin{split}
I &= \gamma^3(v\sqrt{\chi^2-\gamma^{-4}} - \chi + \gamma^{-2})\\
&= \gamma\bigg(1-\frac{1}{2}\chi\bigg) + {\cal O}(\gamma^{-1}).
\end{split}
\end{equation}

Since $\zeta = x^3 - v x^0$, the (angular) frequency $\omega_{\rm BI}$ of the
electrostatic oscillations measured in the lab frame is
\begin{equation}
\label{omega_definition}
\omega_{\rm BI} = \frac{2\pi v}{\lambda} = \frac{2\pi}{\lambda} + {\cal
  O}(\gamma^{-2})  
\end{equation}
and (\ref{lambda_integral_approx}, \ref{definition_I},
\ref{omega_definition}) yield
\begin{equation}
\label{omega_approx}
\omega_{\rm BI} \approx \omega_{\rm AP} \bigg[1 -
\bigg(\frac{\kappa m \omega_p c}{2q}\bigg)^2 \gamma \bigg] 
\end{equation}
where the speed of light $c$ has been restored and $\omega_\text{AP}$
is the (angular) frequency of electrostatic oscillations of a cold
Maxwell plasma for $\gamma\gg 1$~\cite{akhiezer:1956},
\begin{equation}
\omega_{\rm AP} = \frac{\pi}{2\sqrt{2\gamma}} \omega_p.
\end{equation}
%{\bf Discuss value of correction for $\kappa = 10^{-18}$ and $\kappa=
%  10^{-22}$.}
\section{Conclusion}
%%%%%%%%
In practice, a typical laser-plasma experiment currently  has $\gamma \sim 10 -
100$ and $\omega_p \sim 10^{14}~\text{rad}/\text{s}$; using $\kappa\sim
10^{-18}$ it follows $(\omega_\text{BI}/\omega_\text{AP} - 1) \sim
10^{-13} - 10^{-12}$. Although this value is currently out of reach of
terrestrial experiments, we suggest that it may be possible to investigate the
ramifications of (\ref{Emax_BI}, \ref{lower_bound_lambda},
\ref{omega_approx}) in the environment of astrophysical objects such
as quasars and magnetars. In any case, a number of significant
items have been omitted from the above calculation, such as the
finite temperature of the plasma and ion motion. Future work will
include more thorough evaluation of the consequences of non-zero $\kappa$ in the
context of more comprehensive plasma models.  
\section*{Acknowledgements}
%%%%%%%%%
We thank A Noble and RW Tucker for useful discussions, and thank the
Cockcroft Institute of Accelerator Science and Technology for
support. DAB is grateful for support provided by the ALPHA-X
project. RMGMT is supported by the STFC Accelerator Science and Technology
Centre (ASTEC).
%%%%%%

%%%
\end{document}